\documentclass[twocolumn,showpacs,preprintnumbers,amsmath,amssymb]{revtex4}
\usepackage{graphicx,subfigure,subeqnarray,fancyhdr,amsmath,epstopdf,multirow,color, mathrsfs, amssymb}
\def\De{{\rm De}}
\newcommand\y{\mathbf{e}_y}
\newcommand\x{\mathbf{e}_x}
\newcommand\z{\mathbf{e}_z}
\newcommand\sh{\dot{\boldsymbol{\gamma}}}
\newcommand\st{\boldsymbol{\tau}}
\def\Sp{{\rm Sp}}
\newcommand\uvec[1]{\mathbf{\hat{{#1}}}}
\newcommand\I{\boldsymbol{\rm{I}}}
\def\NN{{\rm NN}}
\def\N{{\rm N}}\def\u{{\bf u}}
\def\R{{\rm Re}}

\begin{document}
\title{Enhanced active swimming in viscoelastic fluids}
\author{Emily E. Riley}
\author{Eric Lauga}
\affiliation{Department of Applied Mathematics and Theoretical Physics, 
University of Cambridge, Cambridge CB3 0WA, United Kingdom.}
\begin{abstract}
Swimming microorganisms often self propel in fluids with complex rheology. 
While past theoretical work  indicates that fluid viscoelasticity should hinder their locomotion,  recent experiments on waving swimmers  suggest a  possible non-Newtonian  enhancement of locomotion. 
We suggest a physical mechanism, based on fluid-structure interaction,  leading to swimming in a viscoelastic fluid at a  higher speed than in a Newtonian one. 
Using Taylor's two-dimensional swimming sheet model, we solve for the shape of an active swimmer as a  balance between the external fluid stresses,  the internal  driving moments, and the passive elastic resistance. 
We show that this dynamic balance leads to a generic transition from hindered rigid swimming to enhanced flexible locomotion. The results are physically interpreted as due to a viscoelastic suction increasing the swimming amplitude in a non-Newtonian fluid and overcoming viscoelastic damping.
\end{abstract}

\maketitle
%%%%%%%%%%%%%%%%%%

\section{Introduction}
Active locomotion allows many types of motile cells to adapt to their environment and ensure their  survival~\cite{braybook}. 
One type of oft-studied locomotion is the flagella- or cilia-based  swimming of microorganisms~\cite{childress81}.  
In many instances,  their locomotion  occurs through biological or environmental fluids containing proteins and other polymers which display   elastic, and non-Newtonian, characteristics. 
Important examples include mucus transport by  lung  cilia~\cite{Liron1988}, nematodes travelling though soil~\cite{Wallace1967}, bacteria in their host's tissue~\cite{Suerbaum2002}, and spermatozoa swimming though cervical mammalian mucus~\cite{Pacey2006}.

The  majority of  work  on small-scale swimming  has focused on swimmers self-propelling in Newtonian fluids. Recently, a few  experimental and theoretical studies have addressed the role of non-Newtonian stresses in the fluid, 
with  conflicting conclusions as to their impact on locomotion. On the one hand, measurements with Boger fluids show enhanced propulsion  of a flapping flexible swimmer~\cite{Zenit2013} and of a  
cylindrical swimming sheet~\cite{Kudrolli2013}. While a rotating helix in a Boger fluid displayed decreased swimming at small amplitude, it underwent a modest enhancement at larger helical amplitudes~\cite{spagnolie2013}. 
In contrast, the locomotion of the nematode \emph{C. elegans} was shown to be systematically hindered in a Boger fluid~\cite{Arratia2011}, and similarly  locomotion  in shear-thinning fluids showed a 
systematic decrease~\cite{Kudrolli2013}.  
Additionally recent computational studies on \emph{C.~elegans} showed that both flexibility and back-front asymmetry in stresses are required for enhanced propulsion~\cite{Becca2014}.

In parallel to experiments, analytic studies of locomotion using a prescribed, small-amplitude flagellar waveform  predicted a systematic decrease in swimming speed in viscoelastic fluids~\cite{Lauga2007,Wolgemuth2007,Fu2009}. 
Enhanced swimming has been predicted to occur as either due to  end effects and stress singularities~\cite{Shelley2010} or for large-amplitude swimming~\cite{spagnolie2013}.

In this work we propose a dynamic mechanism for swimming enhancement in a viscoelastic fluid. 
Instead of prescribing the shape of the flagellar deformation we solve for the waveform of the swimmer as a  balance between the fluid stresses and the internal  driving and resisting forces, 
 similarly to work  on actuated finite-size filaments~\cite{Wolgemuth2007} and synthetic swimmers~\cite{Curtis2013}. 
We then show that this dynamic balance  leads to a generic transition from hindered rigid swimming to enhanced flexible locomotion.

%%%%%%%%%%%%%%%%%%
\section{Waving motion in viscoelastic fluids}

In order to model active swimming, we consider Taylor's  two-dimensional waving model \cite{Taylor1951}, which is suitable for the physical description of  
beating eukaryotic  flagella and  cilia~\cite{childress81,Lauga2007}. The extension to the case of three-dimensional filaments is presented in the appendix, with similar results.

\begin{figure}
\centering
\includegraphics[width=0.49\textwidth]{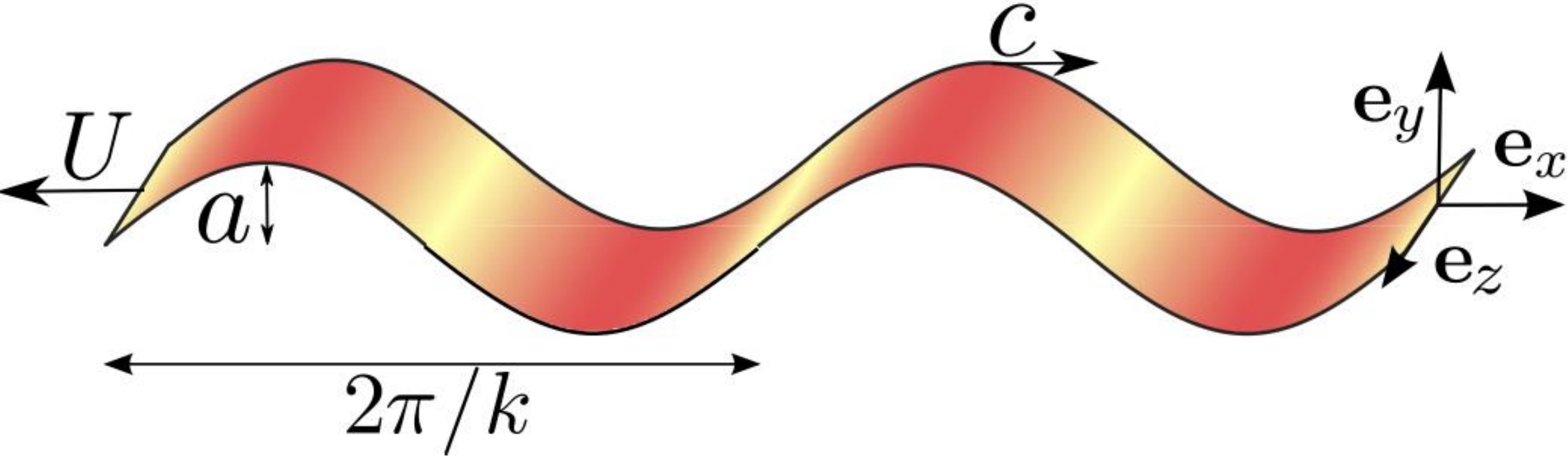}
\caption{A two-dimensional flexible sheet deforming as a traveling wave with  amplitude $a$, wave number $k$, wave speed $c$, and frequency $\omega = k c$ resulting in its swimming at speed $U$.}
\label{sheet}
\end{figure}

Consider an infinite, two-dimensional  sheet of negligible thickness, as illustrated in fig.~\ref{sheet}. 
It is embedded in a viscous fluid, and due to internal actuation is made to deform its shape as a traveling wave of frequency $\omega$, amplitude $a$, wave number $k$, and wave speed $c=\omega/k$, and self-propels as a result with speed $U$ in the opposite direction. 
At low Reynolds number, which is the relevant limit for microorganisms,  the flow around the sheet is described  by the incompressible Cauchy equations, $\nabla\cdot\st=\nabla p$, $\nabla\cdot\mathbf{u}=0$, with $p$ the pressure, $\mathbf{u}$ the velocity field, and $\st$ the deviatoric stress. 
We use a stream function $\psi$ such that $u_x=\partial\psi/\partial y$ and $u_y=-\partial\psi/\partial x$, in order to enforce incompressibility.

Viscoelastic effects in the fluid are modelled using the classical  Oldroyd-B evolution equation for $\st$.  That model, which can be rigorously  derived from  a dilute solution of  infinitely-extensible  elastic dumbbells in a Newtonian solvent~\cite{PhanBook},    captures many features of Boger (elastic, constant viscosity) fluids such as those used in experiments on propulsion \cite{Zenit2013, Kudrolli2013, Breuer2011, Arratia2011}. In the Oldroyd-B model, the deviatoric stress $\st$ is written as a sum, $\st=\st_s+\st_p$, of a Newtonian solvent, $\st_s$,  with viscosity $\eta_s$, and a polymeric stress, $\st_p$,  satisfying a Maxwell model with relaxation time $\lambda$ and viscosity $\eta_p$ (and thus elastic modulus $G\equiv\eta_p/\lambda$).  The total deviatoric stress,  $\st$, satisfies  the Oldroyd-B constitutive equation,
\begin{equation}\label{oldB}
\st+\lambda\stackrel{\triangledown}{\st}=\eta\sh+\eta_s\lambda\stackrel{\triangledown}{\sh},
\end{equation}
where the total viscosity is $\eta=\eta_s+\eta_p$ and where $\sh$ is the shear rate tensor,  $\sh=\nabla \u + \nabla \u^T$. In eq.~\eqref{oldB}, we 
used $ \stackrel{\triangledown}{\mathbf{A}} = {\partial\mathbf{A}}/{\partial t}+\mathbf{u}\cdot\nabla\mathbf{A}-(\nabla\mathbf{u}^T\cdot\mathbf{A}+\mathbf{A}\cdot\nabla\mathbf{u})$, 
to denote the upper convected derivative  for any tensor $\bf A$. An important factor in the derivations below will be $\beta~=~\eta_s/\eta <1$, ratio of solvent to total viscosity. The relative importance of viscoelasticity  is quantified by the    Deborah number, $\De=\lambda\omega$,  ratio of the relaxation time of the polymers  to the relevant  time-scale of the waving motion,   $\omega^{-1}$.

%%%%%%%%%%%%%%%%%%
\section{Swimming speed}

Assuming that the waveform of the sheet is known, we first solve for the external fluid dynamics. 
The height of the sheet is written as $y(x,t) = \epsilon y_1(x,t) + \epsilon^2 y_2(x,t)+\dots$, where  $\epsilon \ll 1$  denotes the dimensionless waving amplitude. The leading-order shape, $y_1$, is decomposed as
\begin{equation}
y_1(x,t)=\R\left[\sum_{n\geq1} a^{(n)}e^{in(kx-\omega t)}\right],
\label{y}
\end{equation}
where $\R$ denotes the real part, and $a^{(n)}$ is the amplitude of the $n$th Fourier mode. Using the Fourier notation $ {W}=\R\left[\sum_{n\geq1}\tilde{W}^{(n)}e^{-in\omega t}\right] $ to describe the $n$th mode  $\tilde{W}^{(n)}$ of a time-periodic function $W$, we thus have $\tilde{y}_1^{(n)}=a^{(n)}e^{inkx}$.

Denoting by  $a^{(n)}_{\NN}$ the sheet amplitude in a non-Newtonian  (Oldroyd-B) fluid and by  $a^{(n)}_{\N}$  the Newtonian one, we can solve for the external fluid dynamics asymptotically in powers of $\epsilon$ following the work in refs.~\cite{Lauga2007, GwynnThesis}. Swimming is obtained at order  $\epsilon^2$, at dimensional speeds in the non-Newtonian (NN) case given  by
\begin{equation}
 U_{\NN}=\tfrac12\sum\limits_{n=1}^{\infty}n^2\omega k
 \big|a^{(n)}_{\NN}\big|
 ^2\left(\frac{1+\beta n^2\De^2}{1+n^2\De^2}\right),
\label{NNSS}
\end{equation}
while in the Newtonian (N) limit we obtain
\begin{equation}
U_{\N}=\tfrac12\sum\limits_{n=1}^{\infty}n^2\omega k\big|a^{(n)}_{\N}\big|^2.
\label{NSS}
\end{equation}
%in the Newtonian (N) limit. 

If the swimmer has identical shape in both fluids, {\it i.e.}~$a^{(n)}_{\NN}=a^{(n)}_{\N}$,  comparing  eqs.~\eqref{NNSS} and \eqref{NSS} shows that we always have $U_{\NN}<U_{\N}$ since $\beta<1$. 
In order to obtain an enhancement of the swimming speed in a viscoelastic fluid, a physical mechanism must thus exist to increase $\big|a^{(n)}_{\NN}\big| $ beyond $\big|a^{(n)}_{\N}\big|$. 
As we show below, solving for the swimmer amplitude by enforcing the correct dynamic balance allows us to obtain a nontrivial dependence of $a^{(n)}_{\NN}$ on the Deborah number, and enhancement.  
As both eqs.~\eqref{NNSS} and \eqref{NSS} are quadratic in the amplitudes $a^{(n)}$, we note that we only need derive the first order shape dynamics.

%%%%%%%%%%%%%%%%%%
\section{Dynamic balance of active swimmer}

Within a beating eukaryotic flagellum there are three forces to consider.  Firstly, the internal driving due to the spatio-temporal actuation from   molecular motors~\cite{Brokaw1989}.  
We model this internal forcing, classically,  as due to a time-varying distribution of active bending moments per unit length,  $F(x,t)$ \cite{Camalet2000}. 
Balancing this actuation are two resisting forces, the external hydrodynamics stresses (pressure and viscous stresses) and the internal solid mechanics resistance  
(elastic cost to be bent away from a preferred, flat state)~\cite{Evans2011}. 
Note that any potential internal dissipation is neglected compared to dissipation in the outside fluid.   
Denoting the bending stiffness of the sheet $\kappa$, normal force balance at leading order in the amplitude of the sheet deformation is written as 
\begin{equation}
-\kappa\nabla^4y+\uvec{n}\cdot\boldsymbol{\sigma}\cdot\uvec{n}|_S=\nabla^2F,
\label{fbal}
\end{equation}
where $\uvec{n}$ is the unit normal to the sheet at leading order and $\boldsymbol{\sigma}$ the hydrodynamic stress tensor. 

In order to determine the hydrodynamic stress, we consider the   constitutive equation, eq.~\eqref{oldB},  at leading order
\begin{equation}\label{order1}
\left(1+\lambda \frac{\partial}{\partial t}\right)\st_1=\eta \left(1+\beta\lambda \frac{\partial}{\partial t}\right)\sh_1,
\end{equation}
where we have expanded each quantity in powers of 
$\epsilon \ll1$, $\st = \epsilon\st_1+\dots$; $\sh=\epsilon\sh_1+\dots$, etc.
Writing eq.~\eqref{order1} using  Fourier notation we have
\begin{equation}
\tilde{\st}_1^{(n)}=\frac{\eta-in\lambda\omega\eta_s}{1-in\lambda\omega}\tilde{\sh}_1^{(n)} = \frac{1-in\De\beta}{1-in\De}\eta\tilde{\sh}_1^{(n)}.
\label{old1st}
\end{equation}
 The first order Stokes equation similarly reduces to
\begin{equation}\label{8}
 \eta\nabla\cdot\tilde{\sh}_1^{(n)}=\frac{1-in\De}{1-in\De\beta}\nabla \tilde{p}_1^{(n)},
\end{equation}
at leading order in $\epsilon$.
The pressure is eliminated from the above by taking its curl, leaving the biharmonic equation for the stream function, $\nabla^4 \tilde\psi_1^{(n)} = 0$. This is solved in Fourier space to obtain the first order stream function as
\begin{equation}
 \psi_1=\R\left[\sum\limits_{n=1}^{\infty}\frac{\omega}{k}a_{\NN}^{(n)}(1+nky)e^{-nky}e^{in(kx-\omega t)}\right].
\label{psi1}
\end{equation}
Notably, the flow described by eq.~\eqref{psi1} is the same as the Newtonian solution, hence viscoelasticity does not modify the flow induced by the swimmer at leading order. However, as we see below, it does impact the stress distribution. 
In order to determine the pressure, we have to integrate eq.~\eqref{8} using eq.~\eqref{psi1}
leading to 
\begin{equation}
\tilde{p}_1^{(n)} =-2\eta\omega k \left(\frac{1-in\De\beta}{1-in\De}\right)in^2a_{\NN}^{(n)}e^{-nky}e^{inkx}.
\label{pressure}
\end{equation}
The hydrodynamic  stress tensor, $\boldsymbol{\sigma}$, is given by $  \boldsymbol{\sigma}=- p\I+\st $, 
which, at leading order, reduces  in Fourier space to
\begin{equation}
  \tilde{\boldsymbol{\sigma}}_1^{(n)}=-\tilde{p}_1^{(n)}\I+\frac{1-in\De\beta}{1-in\De}\eta\tilde{\sh}_1^{(n)}\,,
\label{sigma}
\end{equation}
for each Fourier mode $n$.

\begin{figure*}[hbt]
\centering
\subfigure{
  \includegraphics[height=0.3\linewidth]{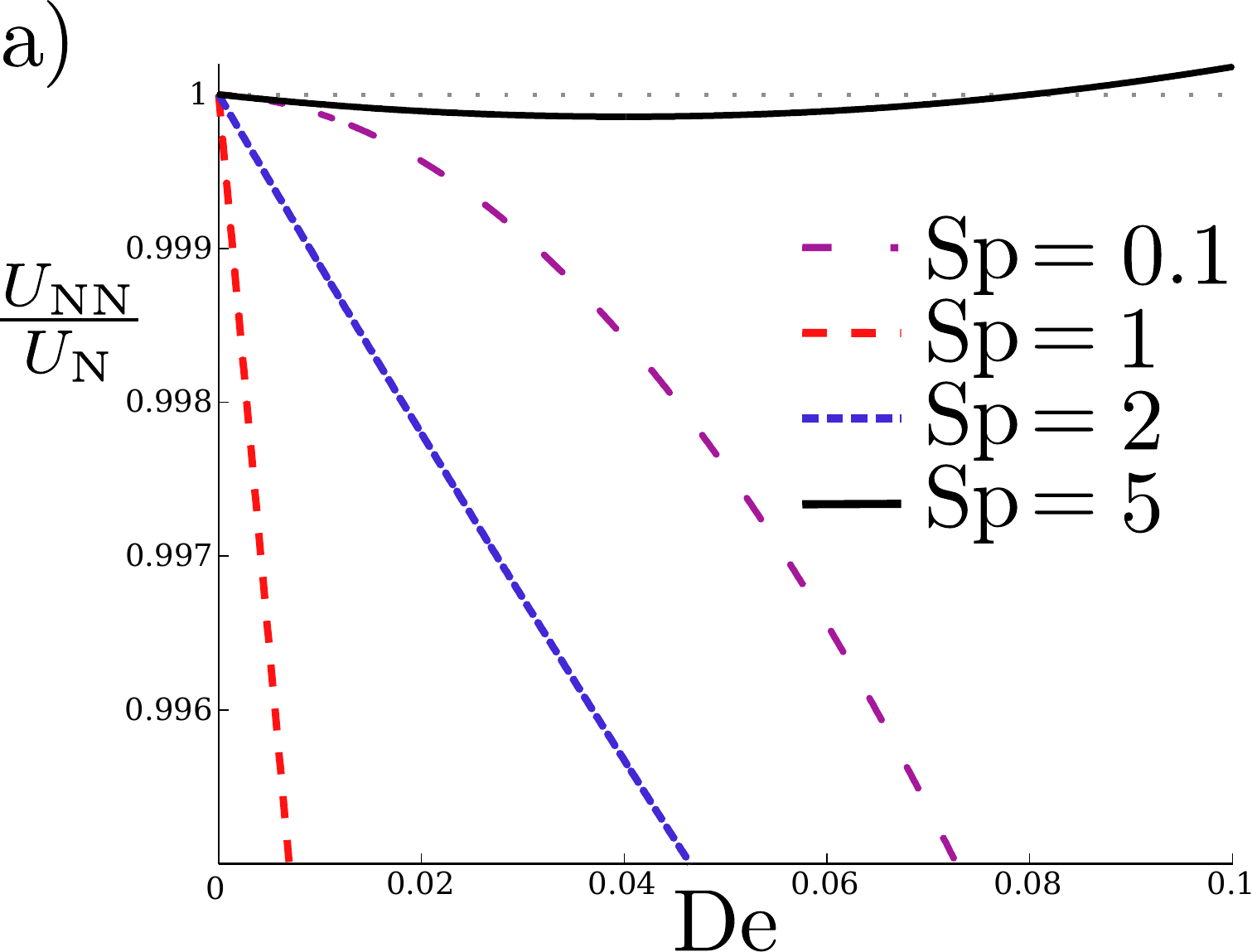}
  \label{plotdeforspa}
} \quad\quad
\subfigure{
  \includegraphics[height=0.3\linewidth]{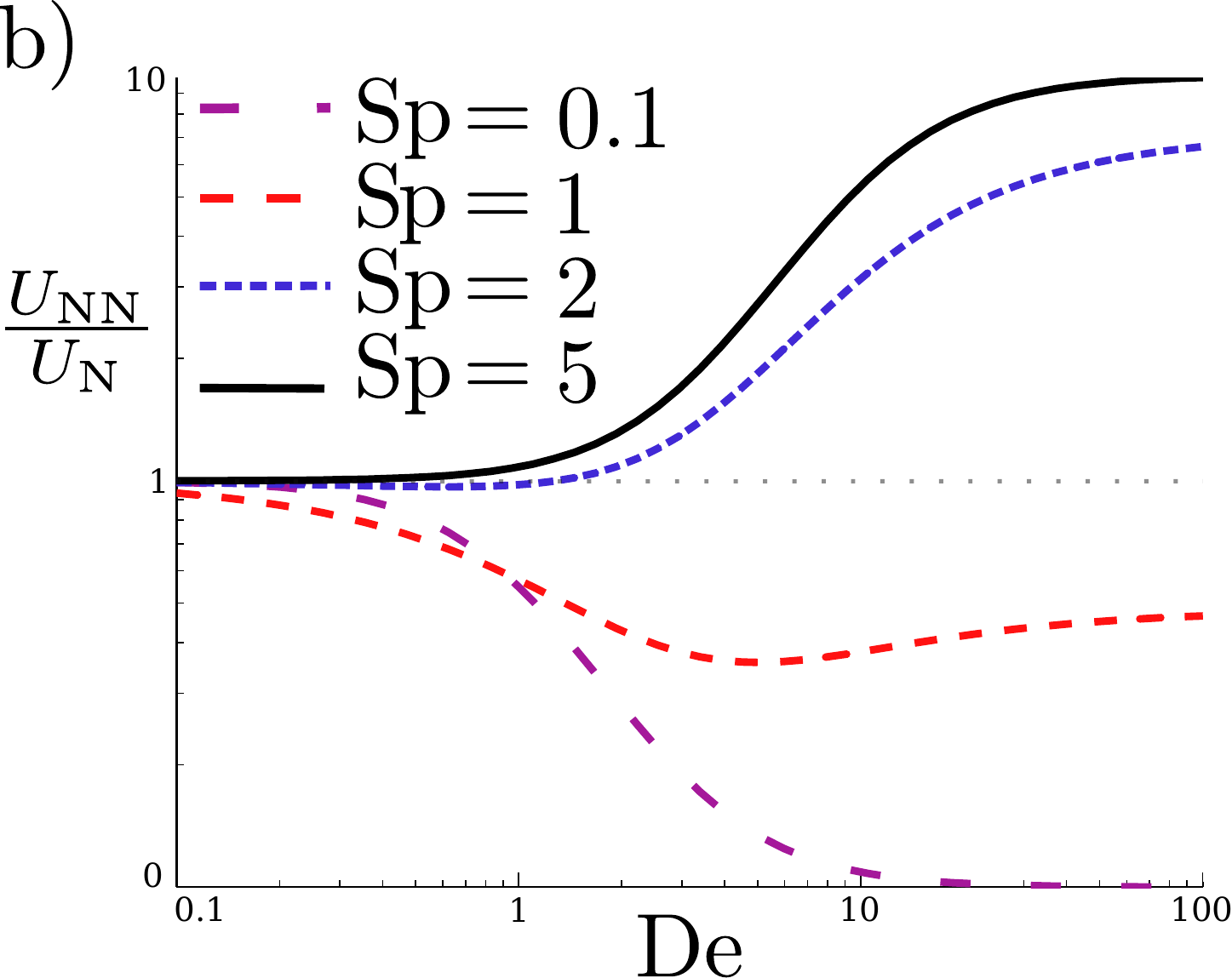}
  \label{plotdeforspb}
} 
\caption{Ratio of non-Newtonian to Newtonian swimming speed, $U_{\NN}/U_{\N}$, as a function of the Deborah number, $\De$,  for four different values of the Sperm numbers ($\beta=0.1$ corresponding to a critical value of  $\Sp \gtrsim 1.16$ for enhanced swimming, eq.~\eqref{splimit}).}
\label{plotdeforsp}
\end{figure*}

With the determination of the fluid stress,  eq.~\eqref{fbal} can be written in Fourier components as 
\begin{equation}
 -\kappa\frac{\partial^4 \tilde{y}^{(n)}_1}{\partial x^4}-\frac{\partial^2 \tilde{F}^{(n)}_1}{\partial x^2}=\tilde{p}_1^{(n)}+2\eta\left(\frac{1-in\De\beta}{1-in\De}\right)\frac{\partial^2\tilde{\psi}_1^{(n)}}{\partial y\partial x}\Bigg|_S
\label{elasto}
\end{equation}
Writing   $
 F_1=\R\left[\sum_{n\geq 1}f^{(n)}e^{in(kx-\omega t)}\right]$ to describe the first order contribution to the active bending moment, we can  determine  the leading-order dynamic  response  of the  sheet amplitude   and obtain
\begin{equation}\label{ampNN}
 a_{\NN}^{(n)}=\frac{-k^2f^{(n)}}{-\kappa n^2 k^4+2\eta\omega ik\displaystyle\left(\frac{1-in\De\beta}{1-in\De}\right)}\cdot
\end{equation}
As can be seen in eq.~\eqref{ampNN}, the value of the Deborah number impacts the sheet amplitude, and thus the swimmer waveform is modified  by a change in the surrounding fluid.

Inputting the linear waveform amplitude, eq.~\eqref{ampNN}, into the quadratic swimming speed, eq.~\eqref{NNSS}, we finally obtain the non-Newtonian swimming speed as
\begin{align}
 U_{\NN}=&\frac{\omega}{2k}\sum\limits_{n=1}^{\infty}\left[\frac{n^2|f^{(n)}|^2}{\kappa^2k^2}\left(\frac{1+n^2\beta\De^2}{1+n^2\De^2}\right)\times\nonumber\right.\\
 &\displaystyle\left.\frac{1}{n^4+4\Sp^6\left(\frac{1+n^2\beta^2\De^2}{1+n^2\De^2}\right)+4n^3\Sp^3\left(\frac{\De(1-\beta)}{1+n^2\De^2}\right)}\right],
\label{NNSSa}
\end{align}
where we have defined the (two-dimensional) Sperm number,
$ \Sp=({\eta\omega}/{\kappa k^3})^{1/3}$, which quantifies the dimensionless ratio of fluid to bending stresses~\cite{Wiggins1998}.  
If $\Sp\ll1$, the dominant balance is between activity and elasticity, and the flagellum waveform is not affected by fluid stresses -- this is  the stiff (s) limit. 
In contrast, when  $\Sp\gg1$, fluid effects balance the active bending and the  waveform changes with the properties of the fluid -- this is the floppy (f) limit.

Simplifying the analysis to focus on the single $n=1$ mode (reducing notation to $f^{(n)}\equiv f$), we have non-Newtonian swimming at speed
\begin{align}
 U_{\NN}=&\frac{|f|^2}{2\kappa^2k^2}\frac{\omega}{k}\times\nonumber\\&\frac{\left(1+\beta\De^2\right)}{1+\De^2+4\Sp^6\left(1+\beta^2\De^2\right)+4\Sp^3\left[\De(1-\beta)\right]},
\label{NNSSA1}
\end{align}
while the Newtonian limit is found by setting   $\De=0$ in eq.~\eqref{NNSSA1}. The non-Newtonian  to  Newtonian swimming speed ratio,  $R={U_{\NN}}/{U_{\N}}$,  is thus given by
\begin{align}
R=\frac{(1+4\Sp^6)(1+\beta\De^2)}{1+\De^2+4\Sp^3\De(1-\beta)+4\Sp^6(1+\beta^2\De^2)},
\label{ratiopb}
\end{align}
which is the main result of this paper.

%%%%%%%%%%%%%%%%%%
\section{Enhanced Locomotion} 

In order to derive the conditions under which swimming enhancement is possible, we need to understand when  the function $R(\beta, \De,\Sp)$ can be above one. Let us first consider some relevant physical limits. In the stiff limit, $\Sp\ll 1$, eq.~\eqref{ratiopb} simplifies to the fixed-amplitude result~\cite{Lauga2007}
\begin{equation}
 R=\frac{1+\beta\De^2}{1+\De^2}\cdot
\end{equation}
In that limit, the  swimming speed ratio decreases monotonically with increasing Deborah number
 to the asymptotic value $U_{\NN}=(\eta_s/\eta)U_{\N}$  for $\De \gg1$.

In the opposite floppy limit,  $\Sp\gg 1$, the flagellum shape is highly sensitive to changes in the hydrodynamic stress and the speed ratio, eq.~\eqref{ratiopb},   reduces to
\begin{equation}
 R=\frac{1+\beta\De^2}{1+\beta^2\De^2}\cdot
\label{Rspinf}
\end{equation}
Here, we obtain a systematic monotonic  increase of the swimming speed with Deborah numbers, up to an asymptotic value  $R=1/\beta$ obtained when   $\De\gg1$. 

Our model  points therefore to a transition from hindered to enhanced swimming when the flagellum is sufficiently flexible. To get further insight, let us look at small deviations from the  Newtonian limit ($\De=0$). 
Computing the derivate of $R$ with respect to $\De$ we get 
${\partial R}/{\partial\De}|_{\De=0}={4(\beta-1)\Sp^3}/{(1+4\Sp^6)}$, 
which is always negative. Consequently, a small amount of viscoelasticity ($\De \ll 1$) will always start by decreasing the swimming speed. In contrast, in the  infinite Deborah number limit, the 
swimming speed ratio becomes
\begin{equation}
 R(\De\gg1 )=\frac{\beta+4\beta\Sp^6}{1+4\beta^2\Sp^6}\cdot
\end{equation}
A transition from hindered ($R<1$) to enhanced propulsion ($R>1$) in a non-Newtonian fluid occurs thus when
\begin{equation}
\Sp^3>\frac{1}{2\sqrt{\beta}}\cdot
\label{splimit}
\end{equation}
The result in eq.~\eqref{splimit} indicates therefore  a transition in swimmer flexibility allowing enhancement of the swimming speed. Indeed, the Sperm number scales  inversely proportional to the flagellum bending modulus, and thus for a given fluid, the criterion in eq.~\eqref{splimit} is equivalent to a requirement  for $\kappa$ to be small enough.

Our results are illustrated numerically in fig.~\ref{plotdeforsp} for  $\beta=0.1$. We plot the ratio of the non-Newtonian to Newtonian swimmer speed, $U_{\NN}/U_{\N}$, as a function of the Deborah number for four different values of the Sperm number. The data are shown in fig.~\ref{plotdeforsp}a for small values of $\De$ and  ranging from 0 to 100 in fig.~\ref{plotdeforsp}b.
In all cases, the swimming speed initially decreases with the Deborah number (fig.~\ref{plotdeforsp}a) but when the swimmer is sufficiently flexible, the swimming speed subsequently increases and crosses the threshold $U_{\NN}/U_{\N}=1$ (fig.~\ref{plotdeforsp}b). The criterion  from eq.~\eqref{splimit}  corresponds  to enhancement predicted to occur as soon as $\Sp \gtrsim 1.16$, consistent with the numerical results. Note that our model also allows us to compute the value of the transition Deborah number beyond which enhancement occurs. 
In eq.~\eqref{ratiopb}, one can solve the quadratic equation for $\De$ and $R>1$ is equivalent to 
$
 \De>{4\Sp^3}/({4\beta\Sp^6-1}), 
$ which, as expected, is defined only if the criterion in eq.~\eqref{splimit} is satisfied. 

Beyond swimming kinematics, our model also allows us to compute  swimming  energetics and  efficiency. Following ref.~\cite{Lauga2007} and the derivations above, we can calculate  the power expanded  by the swimmer against the fluid, $\dot{\cal W}_{\NN}$. Defining the swimming efficiency, classically, as ${\cal E} = \eta U^2 / \dot {\cal W}$,  the ratio between the non-Newtonian efficiency and that in  a Newtonian fluid with the same viscosity ($\eta$) is exactly given by the swimming speed ratio $R$ from Eq.~\eqref{ratiopb}.  The conditions for enhanced swimming correspond thus to those required for enhanced efficiency.

%%%%%%%%%%%%%%%%%%
\section{Illustration of the waveform}

\begin{figure*}
\centering
\includegraphics[width=0.8\textwidth]{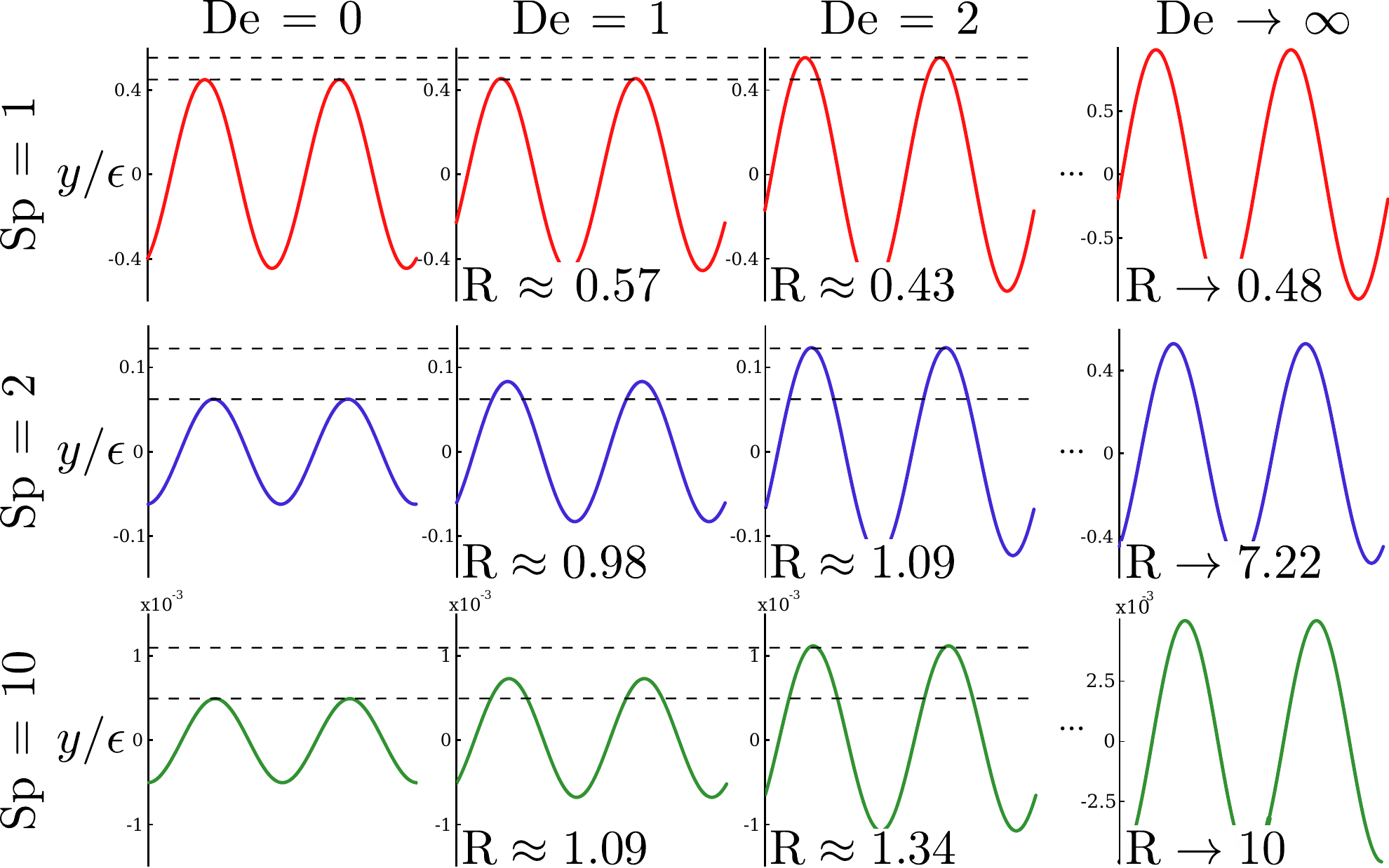}
\caption{Swimming  waveforms under a linear response for  $\Sp=1,2,$ and 10, as a function of the Deborah number, $\De$. In each plot, the value of the swimming   ratio, $R=U_{\NN}/U_{\N}$, is indicated. As in fig.~\ref{plotdeforsp} $\beta=0.1$ and    $\Sp \gtrsim 1.16$ is required  for enhancement. An increase in $\De$  leads to an increase in waving amplitude which, when sufficiently large, leads to enhanced swimming.}
\label{spwaves}
\end{figure*}

We further illustrate the impact of non-Newtonian stresses by displaying the swimming waveform in the case of an internal sinusoidal forcing. 
We thus prescribe   $f^{(1)}=fi$ and $f^{(n)}=0$ for $n >1$, and compute the leading-order waveform. Under the assumption of linear response, the shape remains sinusoidal with a different phase and amplitude. The results are illustrated in fig.~\ref{spwaves} for three Sperm numbers (1, 2 and 10) and four Deborah numbers (0, 1, 2, and $\infty$). Superimposed on the shapes  are the values of $R$, ratio of the non-Newtonian to Newtonian swimming speeds.  

The results in fig.~\ref{spwaves} show  the expected decrease in waving amplitude that accompanies an increase in Sperm number  but, more importantly,   the systematic increase in amplitude with an increase of viscoelasticity $\De$. The waving amplitudes can be computed analytically and we obtain the Newtonian result as 
$ A_{\N}^2 = {1}/{{(1+4\Sp^6)}}$,
which explains the decrease of waving amplitude with Sperm number. 
In the  non-Newtonian case we have a waving amplitude given by
\begin{equation}
 A_{\NN}^2 = {\frac{1+\De^2}{1+\De^2+4\Sp^3\De(1-\beta)+4\Sp^6(1+\beta^2\De^2)}}\cdot
\end{equation}
The critical Deborah number for which $A_{\NN}>A_{\N}$ is  then found to be 
$ \De>{1}/{\Sp^3(1+\beta)}, $
which is always satisfied. Hence the presence of viscoelastic stresses leads to a systematic increase of the waving amplitude of the swimmer. This provides    a physical interpretation for the  swimming enhancement seen in fig.~\ref{plotdeforsp}: if the viscoelastic amplitude increase is large enough, it is able to   compensate for the non-Newtonian damping term from eq.~\eqref{NNSS}, leading to faster swimming, $U_{\NN} > U_{\N}$.

%%%%%%%%%%%%%%%%%%
\section{Discussion} 
In this paper,  we have proposed a physical mechanism for enhanced locomotion in a viscoelastic fluid. It does not require  transient or end  effects but 
instead arises naturally due to the equations of active elastohydrodynamics applied to locomotion. 

Our results can be rationalised   by focusing on the two different stiff (s) and floppy (f)  limits, and comparing the non-Newtonian swimming speed for  $\De \gg 1$ to the Newtonian one ($\De=0$). 
In the stiff regime fluid forces are negligible, and 
the dynamic balance in  eq.~\eqref{elasto} reduces to one between  bending  resistance and active stresses. Considering only the typical magnitudes of $a$ and $f$, we then get  $a^{(s)} \sim f/\kappa k^2 $  for both Newtonian and non-Newtonian.   
 The swimming speeds scale then as $U_{\N}^{(s)}\sim\omega f^2/\kappa^2k^3$ and, for large $\De$,  $U_{\NN}^{(s)}\sim\beta\omega f^2/\kappa^2k^3$, leading to  
$R^{(s)}=\beta < 1$. In contrast, in the  floppy regime,  elastic  forces are negligible compared to  fluid stresses, and the dynamic balance in eq.~\eqref{elasto} reduces to  one between the fluid resistance of the filament and the active stresses, with 
$a^{(f)}_{\N} \sim kf/\eta\omega $
and $a^{(f)}_{\NN} \sim kf/\eta\omega\beta $.  
The swimming  speeds in this case are given by $U_{\N}^{(f)}\sim k^3f^2/\eta^2\omega$ and $U_{\NN}^{(f)}\sim k^3f^2/\beta\eta^2\omega$, leading to $R^{(f)}=1/\beta > 1$, and enhanced swimming. 
%check scaling section

Physically, we have shown  that the transition from hindered to enhanced swimming takes its origin in the systematic increase of the waving amplitude for active swimming in a viscoelastic fluid, which can overcome  
viscoelastic fluid damping  \cite{Lauga2007}. How can this increase in amplitude be intuitively rationalised? We would like to argue that it is a consequence of the change in fluid pressure, and  results from a `viscoelastic suction'.  Indeed, we consider the leading-order pressure in eq.~\eqref{pressure}, and compute  its typical value on the sheet for a {fixed} amplitude $a$, allowing us to isolate the change  in   pressure due to the fluid 
dynamics and not due to the amplitude increase. The ratio between the typical non-Newtonian and Newtonian pressure is then
\begin{equation}
\left[\frac{p_{\NN}(a)}{p_{\N}(a)}\right]^2= {\frac{1+\beta^2\De^2}{1+\De^2}},
\end{equation}
which shows a large pressure reduction (since $\beta<1$) in a viscoelastic fluid. For sufficiently large  Sperm numbers, where fluid stresses  have a relatively larger impact on the waveform, the wave amplitude  increase due to this suction effect is able to overcome the non-Newtonian fluid damping in locomotion and  increase the swimming speed.

To conclude, we note that although the mechanism outlined in this paper was derived in the context of flagellar locomotion, the same physical principle would be at play for higher swimmers exploiting  muscular contractions, and thus could also be relevant to the dynamics  of small multicellular organisms in complex environments.

%%%%%%%%%%%%%%%%%%
\section*{Appendix: infinite filament}
\label{A}
\renewcommand{\theequation}{A.\arabic{equation}} 
 \setcounter{equation}{0}
Having studied a two-dimensional  waving sheet, we  outline how to carry out the calculation  a three-dimensional infinite filament, following 
ref.~\cite{Powers2008}. Consider an infinite periodic filament in an Oldroyd-B fluid waving with small amplitude. The filament is modelled geometrically  as a cylinder, which when straight is parametrised by $s$ along its  
axis, $\phi$ around this axis, and  radius $b$. When small-amplitude waves propagate along ${\bf e}_x$, the surface of the cylinder is described by $\uvec{r}=[h(s,t)+b\cos\phi]\y+b\sin\phi\z+s\x$ and the height of the filament away from 
its centreline position is 
\begin{equation}
h(s,t)=\epsilon\R\left[a_{\NN}e^{i(ks-\omega t)}\right], 
\end{equation}
where $a$ is the amplitude, and $h(s,t)$ is analogous to $y(x,t)$ in the above when $n=1$. 
Using the first-order Oldroyd-B equation, in Fourier notation, we recover eq.~\eqref{old1st}. 
We can then consider the Stokes equation at first order,
\begin{equation}
 \nabla^2\tilde{\mathbf{u}}_1=\eta\left(\frac{1-i\beta\De}{1-i\De}\right)\nabla\tilde{p}_1,
\label{stokes}
\end{equation}
where the first order boundary conditions for each mode are given by $\tilde{\mathbf{u}}_1=i\omega\tilde{h}\y$. The solutions   are best derived using cylindrical polar co-ordinates, where the basis vectors are 
$\x$, $\uvec{r}=\sin\phi\z+\cos\phi\y$ and $\uvec{\theta}=-\sin\phi\y+\cos\phi\z$. %
\begin{table}[t]
{\begin{tabular}{l c}
\hline\hline
\\
$\tilde{p}$ & $-\eta\omega k\tilde{h}i\cos\theta AK_1(kr)$\\
$\tilde{u}_r$&  $ -\tilde{h}\omega i\cos\theta[\alpha ArkK_1(kr)+BK_2(kr)+CK_0(kr)]$\\
$\tilde{u}_{\theta}$ & $ -\tilde{h}\omega i\sin\theta[BK_2(kr)-CK_0(kr)] $ \\
$\tilde{u}_x$ & $-\tilde{h}\omega\cos\theta[\alpha ArkK_0(kr)+(B+C-\alpha A)K_1(kr)] $ \\
$\alpha A$ & $\left\{K_0(kb)+b{k}K_1(kb)\left[{\frac{1}{2}+}\frac{K_0(kb)}{2K_2(kb)}-\frac{K_0(kb)^2}{K_1(kb)^2}\right]\right\}^{-1}$\\ 
$B$ & $-\alpha Ab{k}K_1(kb)/[2K_2(kb)] $\\
$C$ &$[1-b{k}K_1(kb)\alpha A/2]/K_0(kb) $\\
\hline\hline
\end{tabular}}
\caption{Pressure and velocity field for a travelling wave in the $\x$ direction on a cylinder to first-order in $\ln(kb)$, {with $\alpha=(1-i\De)/(1-i\beta\De)$} \cite{Fu2009}.}
\label{filament}
\end{table}
The first-order solutions are shown in table~\ref{filament}~\cite{Fu2009}. These can then be used to compute the second-order swimming speed, leading to the same result as eq.~\eqref{NNSS}. 
The force per unit length acting of the fluid is found by integrating the stress around the circumference and keeping  only lowest-order terms in $\ln(kb)$  the force per unit length perpendicular to the filament is
\begin{equation}
{\mathcal{F}_{vis}=\R\left[-i\omega a_{\NN}\frac{4\pi}{\ln(kb)}\frac{1-i\beta\De}{1-i\De}\eta e^{i(ks-\omega t)}\right]}.
\end{equation}
This viscoelastic force is then balanced by the passive elastic forces of the filament and the internal forcing, 
\begin{equation}
{\tilde{\mathcal{F}}_{vis}= -{2b}\frac{\partial^2\tilde{F_{1}}}{\partial s^2}- \kappa\frac{\partial^4 \tilde{h}}{\partial s^4}}\,,
\end{equation}
where $F_{1}$ is the first-order bending moment per unit length, as above. The non-Newtonian swimming velocity is then finally given by
\begin{align}
&{U_{\NN}=\frac{{2b^2}\left|{f}\right|^2}{{\kappa^2}{k^2}}\frac{\omega}{k}\times}\nonumber\\
&{\frac{\left(1+\beta\De^2\right)}{1+\De^2+{2}\Sp^4\De(1-\beta)+\Sp^8\left(1+\beta^2\De^2\right)}},
\end{align}
where we have defined the perpendicular drag coefficient $\xi_{\perp}=4{\pi}\eta/\ln(kb)$, and the three-dimensional Sperm number, $\Sp=({\xi_{\perp}\omega/\kappa k^4})^{1/4}$. 
The three dimensional result,  eq.~{\eqref{NNSSA1}}, is thus very similar to the two-dimensional case, and the physical mechanism identified in this paper extends naturally to three dimensions.  

%%%%%%%%%%%%%%%%%%
\begin{acknowledgments}
We thank RE Goldstein and TJ Pedley for useful discussions. This work was funded in part by the European Union (CIG grant to EL).
\end{acknowledgments}

\end{document}